\documentstyle[preprint,aps,epsf,floats]{revtex}

\begin{document}

\vskip 1.5 cm

\preprint{\tighten \vbox{\hbox{} }}

\title{Using the value of $\beta$ to help determine $\gamma$ from $B$ decays}
\author{Cheng-Wei Chiang\footnote{E-mail:chengwei@andrew.cmu.edu} and 
Lincoln Wolfenstein\footnote{E-mail:lincoln@cmuhep2.phys.cmu.edu}}

\address{
Department of Physics,
Carnegie Mellon University,
Pittsburgh, Pennsylvania 15213}

\maketitle

{\tighten
\begin{abstract}
It has been pointed out by Gronau and Rosner that the angle $\gamma$
of the unitarity triangle could be determined by combining future
results on $B_s$ and $B_d$ decays to $K \pi$.  Here we show that it is
important to include in the analysis the information on the phase
$\beta$ which will be determined in the near future.  Omitting this
information could lead to an error as large as $8^{\circ}$ in
$\gamma$.
\end{abstract}
}
\vspace{0.2in}

\newpage

A large number of experiments have been proposed to determine the
phase $\gamma=Arg(V_{ub}^*)$ in the CKM matrix
\cite{GW91,LGR94,DH95,FM98,GR98,NR98,N98,BF99}.  Before any of these
experiments is completed it is likely that there will be a good
measurement of $\sin 2\beta$.  In many cases using the value of
$\beta=Arg(V_{td}^*)$ derived from $\sin 2\beta$ can improve possible
determinations of $\gamma$.  We illustrate this for the case of a
recent proposal by Gronau and Rosner \cite{GR00} to determine $\gamma$
by using U-spin symmetry (the exchange of $s$ and $d$ quarks) to
relate the decays $B^0 \to K^+ \, \pi^-$ to $B_s \to K^- \, \pi^+$.
Combining the rate of these decays with the rate for $B^+ \to K^0
\pi^+$ the value of $\gamma$ could be obtained.  We assume throughout
the constraints of the CKM model.

The tree amplitude for $B^0$ ($B_s$) decay is proportional to
$V_{ub}^*V_{ud}$ ($V_{ub}^*V_{us}$).  The penguin amplitude is
dominated by the virtual $t$ quark and is proportional to
$V_{tb}^*V_{ts}$ ($V_{tb}^*V_{td}$).  Their approximation is to assume
that the decay $B^+ \to K^0 \pi^+$ is purely penguin because only the
penguin gives $b \to s \, \bar d \, d$.  We then find for the decay
amplitudes
\begin{mathletters}
\label{amp}
\begin{eqnarray}
\label{amp-a}
A(B^+ \to K^0 \, \pi^+) &=& P, \\
\label{amp-b}
A(B^0 \to K^+ \, \pi^-) &=& T \, e^{i(\delta+\gamma)} + P, \\
\label{amp-c}
A(B_s \to K^- \, \pi^+) &=& 
\frac1{{\tilde \lambda}} T^{\prime} \, e^{i(\delta^{\prime}+\gamma)}
  - P^{\prime} \left| \frac{V_{td}}{V_{ts}} \right| e^{-i \beta};
\end{eqnarray}
\end{mathletters}
where ${\tilde \lambda} \equiv \left|V_{us}/V_{ud}\right| \simeq
0.226$.  $\left| V_{td}/V_{ts} \right|$ is completely determined in
terms of $\beta$, $\gamma$, and $\tilde\lambda$.  The U-spin
approximation is $P^{\prime}=P$, $T^{\prime}=T$, and
$\delta^{\prime}=\delta$.

In Ref.~\cite{GR00} unitarity is used to set
\begin{equation}
V_{tb}^* V_{ti} = -\left( V_{cb}^* V_{ci} + V_{ub}^* V_{ui} \right),
\end{equation}
for $i=\,d,s$.  Thus part of what we have called the penguin is now in
the $V_{ub}^* V_{ui}$ term and combined with the tree; therefore, they
get
\begin{mathletters}
\label{GRamp}
\begin{eqnarray}
\label{GRamp-a}
A(B^+ \to K^0 \, \pi^+) &=& \bar P, \\
\label{GRamp-b}
A(B^0 \to K^+ \, \pi^-) &=&
  \bar T \, e^{i(\bar\delta+\gamma)} + \bar P, \\
\label{GRamp-c}
A(B_s \to K^- \, \pi^+) &=& 
  \frac1{{\tilde \lambda}} \bar T^{\prime} \, e^{i(\bar\delta^{\prime}+\gamma)}
     - {\tilde \lambda} \bar P^{\prime};
\end{eqnarray}
\end{mathletters}
where $\bar\delta$ and $\bar\delta^{\prime}$ are in general different
from $\delta$ and $\delta^{\prime}$ in Eqs.~(\ref{amp}) and the last
term follows since $V_{cd}/V_{cs}=-{\tilde \lambda}$.  They thus
obtain simple results independent of $\beta$.

However, terms of ${\cal O}({\tilde \lambda}^2)$ and with dependence
on both $\beta$ and $\gamma$ have been omitted from the first equation
in (\ref{GRamp}).  Since $\beta$ will be known when this analysis can
be used there is no purpose in eliminating $\beta$.  We instead use
Eqs.~(\ref{amp}) to determine $\gamma$ and the ratio $r \equiv P/T$
from the quantities $R_d$ and $R_s$ defined in Ref.~\cite{GR00}
\footnote{$R_d$ is the ratio of the sum of $B^0$ and ${\bar B}^0$
decays to that of $B^+$ and $B^-$ decays.  $R_s$ is the ratio of the
sum of $B_s$ and ${\bar B}_s$ decays to that of $B^+$ and $B^-$
decays.} for any value of $\beta$.  Typical results are shown in
Figs.~1 and 2 where we fix the sum of $R_d$ and $R_s$ and consider the
limiting case $\delta=\delta^{\prime}=0$.  The results of
Ref.~\cite{GR00} are reproduced in the limit $\beta=0$.

\begin{figure}[t]
\centerline{\epsfysize=8truecm  \epsfbox{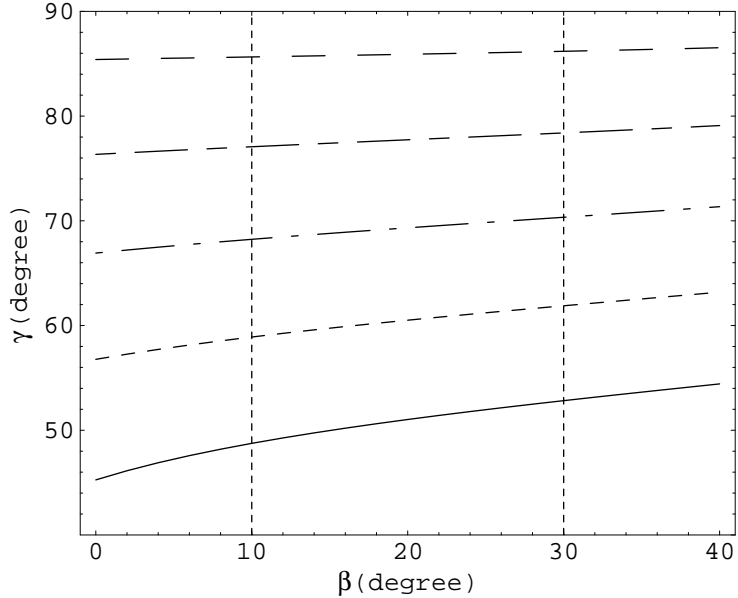} }
\vspace{0.2in}
\tighten{
\caption[]{\it The dependence of $\gamma$ on $\beta$ for $R_d=0.8$,
$R_s=0.78$ (solid line), $R_d=0.85$, $R_s=0.73$ (short dashed line),
$R_d=0.9$, $R_s=0.68$ (dash-dot-dash line), $R_d=0.95$, $R_s=0.63$
(short-long dashed line), and $R_d=1$, $R_s=0.68$ (long dashed line),
respectively.  We assume $\cos\delta=1$ and $r<0$.}}
\end{figure}
\begin{figure}[t]
\centerline{\epsfysize=8truecm  \epsfbox{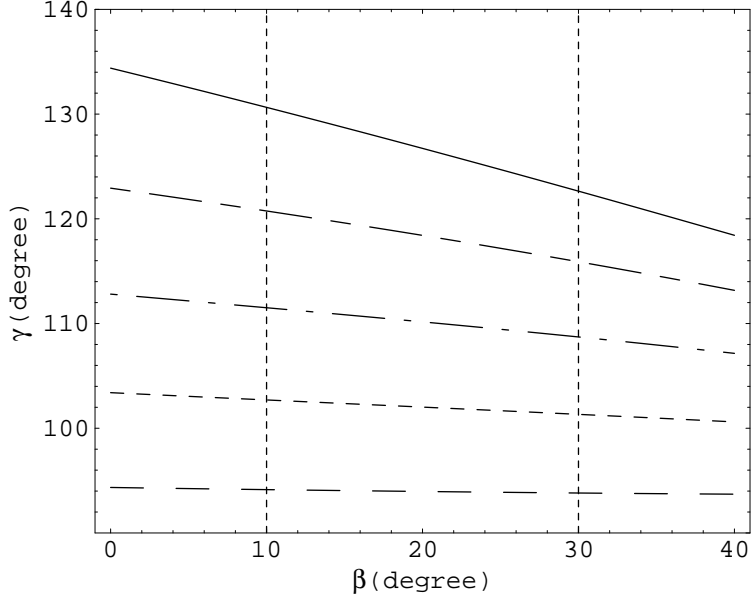} }
\vspace{0.2in}
\tighten{
\caption[]{\it The dependence of $\gamma$ on $\beta$ for $R_d=1.05$,
$R_s=0.53$ (long dashed line), $R_d=1.1$, $R_s=0.48$ (short dashed
line), $R_d=1.15$, $R_s=0.43$ (dash-dot-dash line), $R_d=1.2$,
$R_s=0.38$ (short-long dashed line), and $R_d=1.25$, $R_s=0.33$ (solid
line), respectively.  We assume $\cos\delta=1$ and $r<0$.}}
\end{figure}

It is seen that for values of $\gamma$ in the neighborhood of
$50^{\circ}$ and for $\beta=30^{\circ}$ ($\sin 2\beta=0.87$) the
values of $\gamma$ is shifted from $\sim45^{\circ}$ to
$\sim53^{\circ}$ from the $\beta=0$ approximation.  For values of
$\gamma$ in the neighborhood of $130^{\circ}$ and for
$\beta=20^{\circ}$ the shift is from $\sim134^{\circ}$ to
$\sim127^{\circ}$.  We assume $r<0$ in accordance with the
factorization assumption.

It is instructive to analyze the difference in the two approximations.
The effective interaction can be written as
\begin{equation}
{\cal H}_{eff} = V_{tb}^* V_{tq} \sum_{i=3}^6 \,Q_i
               + V_{ub}^* V_{uq} \sum_{i=1}^2 \,Q_i^{(u)}
               + V_{cb}^* V_{cq} \sum_{i=1}^2 \,Q_i^{(c)},
\end{equation}
where $q=d$ or $s$ and $Q_i$ are the standard operators including the
Wilson coefficients.  We use the approximation that annihilation
diagrams can be neglected so that $B^+ \to K^0 \pi^+$ is due to the
penguin operators $Q_3 \sim Q_6$.  Thus, as assumed in deriving
Eqs.~(\ref{amp}) the terms $P \, (P^{\prime})$ are proportional to
$V_{tb}^* V_{ts}\,(V_{tb}^* V_{td})$.  In Ref.~\cite{GR00} unitarity
is used to set
\begin{equation}
- V_{tb}^* V_{ts} = V_{cb}^* V_{cs} + V_{ub}^* V_{us}
\end{equation}
and then the term proportional to $V_{ub}^* V_{us}$ is just omitted on
the ground that it is smaller by a factor $\lambda^2$.  In the limit
that we neglect the strong phases we can include this term by
replacing Eq.~(\ref{GRamp-a}) by
\begin{equation}
\label{chargeB}
A(B^+ \to K^0 \, \pi^+) = \bar P
  \left[
    1 + {\tilde \lambda}^2 \frac{\sin \beta}{\sin (\beta+\gamma)} e^{i \gamma}
  \right].
\end{equation}
Formally our results reduce to theirs in the limit $\beta=0$.  It is
the amplification of this factor ${\tilde \lambda}^2$ that is
responsible for the difference.

The equations of Ref.~\cite{GR00} for $R_d$ and $R_s$ become equations
for $K\,R_d$ and $K\,R_s$, where
\begin{equation}
\label{K}
K = 1 + 2 {\tilde \lambda}^2 \frac{\sin\beta \cos\gamma}{\sin(\beta+\gamma)}
      + {\tilde \lambda}^4 
        \left( \frac{\sin\beta}{\sin(\beta+\gamma)} \right)^2.
\end{equation}
The same factor $K$ enters for $R_d$ and $R_s$ because both are
defined as ratios to the $B^+$ decay.  Then
\begin{eqnarray}
\label{Ratio}
K\,R_d &=&
  1 + r^2 + 2r \cos\delta \cos\gamma,
  \nonumber \\
K\,R_s &=&
  {\tilde \lambda}^2 + \left(\frac{r}{\tilde \lambda}\right)^2
                     - 2r \cos\delta \cos\gamma.
\end{eqnarray}

The amplification arises from the fact that $r\cos\gamma$ is
proportional to $(K\,R_d-1-r^2)$.  Thus, for example, with values of
$R_d=0.8$ and $R_s=0.78$ (corresponding to $\gamma\sim50^{\circ}$) a
change of $K$ from $1$ to $1.03$ decreases $\left|\, r\cos\gamma
\,\right|$ by about $10\%$.  The value of $r^2$ is proportional to
$\left[ K(R_d+R_s)-(1+\tilde\lambda^2) \right]$.  For our example with
$R_d+R_s=1.58$ a change of $K$ from $1$ to $1.03$ increases $r$ by
about $5\%$.  Thus, a change of $K$ from $1$ to $1.03$ can decrease
$\cos\gamma$ by about $15\%$.

Unfortunately, the difficulty of using this method arises from the
same sensitivity; small errors on $R_d$ and $R_s$ can cause a
significant error on the determined $\gamma$.  As an example, let the
experimental errors be
\begin{equation}
\frac{\Delta R_s}{R_s} = 2 \frac{\Delta R_d}{R_d} \equiv 2 \epsilon.
\end{equation}
For the case shown in Fig. 1 with $\beta=30^{\circ}$ and
$\gamma=53^{\circ}$, a value of $\epsilon=6\%$ corresponds to an
uncertainty of about $24\%$ in $\cos\gamma$, yielding a value
$\gamma=53^{\circ}\pm10^{\circ}$.  For another case in Fig. 2 with
$\beta=18^{\circ}$ and $\gamma=128^{\circ}$ and assuming instead
$\Delta R_s/R_s = 4 \Delta R_d/R_d \equiv 4 \epsilon$, the same value
of $\epsilon$ would correspond to an error of about $23\%$ in
$\cos\gamma$ and $\gamma=128^{\circ}\pm10^{\circ}$.

The accuracy of this method requires including the strong phase
$\delta$.  In principle this can be determined by measuring the
asymmetry between the rates for $B^0$ and $\bar B^0$, which is
proportional to $\sin \gamma \sin \delta$.  To a first approximation,
the quantity that is determined in the method discussed here is
$\cos\gamma\,\cos\delta$.  Assuming $\delta$ is small probably only a
limit on $\sin\gamma\,\sin\delta$ can be achieved.  If
$\cos^2\gamma<1/2$ and $\sin\gamma\,\sin\delta<X$, then the
uncertainty in $\delta$ leads to an error of no more than $0.35\,X^2$ in
$\cos\gamma$.  It should be emphasized that this method depends upon
the assumption that the sign of $r$ is as given by factorization.

The approximation of neglecting contributions from $Q_1$ and $Q_2$
needs to be considered.  The contribution of $Q_i^{(c)}$ can be
included in ${\bar P}$ since in going from Eqs.~(\ref{GRamp-a}) to
(\ref{GRamp-b}) all that is required is that ${\bar P}$ corresponds to
no change in isospin.  As a result the only effect is a correction to
the term proportional to ${\tilde \lambda}^2$ in Eq.~(\ref{chargeB}).
The contributions to $Q_1$ and $Q_2$ are long-distance effects due to
rescattering which mixes processes of different topologies;
calculations of these effects are very model dependent
\cite{FSI,CFMS97,FKNP98}.  If we call $P_u (P_c)$ the amplitudes due
to $Q_1^{(u)}+Q_2^{(u)} (Q_1^{(c)}+Q_2^{(c)})$ then the ${\tilde
\lambda}^2$ terms in Eq.~(\ref{chargeB}) must be multiplied by
$1+(P_u-P_c)/{\bar P}$.  Ciuchini {\it et.~al.} \cite{CFMS97}, who
call $P_c$ the ``charming penguin'', suggest that $P_c/{\bar P}$ could
be of order unity and Falk {\it et.~al.}  \cite{FKNP98} suggest that
$P_u/{\bar P}$ could be large.  However, a recent analysis by Kamal
\cite{K99} suggests that $(P_u-P_c)/{\bar P}$ is probably of order
$0.1$.  As pointed out in these papers, it should be possible in the
future to limit the values of $P_u$ and $P_c$ by detecting decays
where they would make a major contribution.

In conclusion we emphasize that in determining $\gamma$ from future
experiments, optimum use should take into account the value of $\beta$
which will be measured via $\sin 2\beta$ in the near future.  In the
examples we have discussed of $B_d$ ($B_s$) decays to $K \pi$, the
omission of the $\beta$ dependence could lead to an error as large as
$8^{\circ}$ in special cases.  In the longer run it would be valuable
to determine the phase of the penguin amplitude and the phase $2\beta$
of the mixing independently so as to detect new physics contributions.
Here we have limited the discussion to the standard CKM model.

\smallskip

This research is supported by the Department of Energy under Grant
No. DE-FG02-91ER40682.


{\tighten

}

\end{document}